\documentclass[%
 reprint,
 amsmath,amssymb,
 aps,
]{revtex4-1}

\usepackage{graphicx}
\usepackage{dcolumn}
\usepackage{bm}

\begin{document}

\preprint{APS/123-QED}

\title{High pressure effects on non-fluorinated BiS$_2$-based superconductors 
La$_{1-x}$$M$$_x$OBiS$_2$ ($M$ = Ti and Th) }

\author{Y. Fang$^{1,2}$}
\author{D. Yazici$^{2,3}$}
\author{I. Jeon$^{1,2}$}
\author{M. B. Maple$^{1,2,3}$}
\email[Corresponding Author: ]{mbmaple@ucsd.edu}

\affiliation{$^1$Materials Science and Engineering Program, University of California, San Diego, La Jolla, California 92093, USA}
\affiliation{$^2$Center for Advanced Nanoscience, University of California, San Diego, La Jolla, California 92093, USA}
\affiliation{$^3$Department of Physics, University of California, San Diego, La Jolla, California 92093, USA}

\begin{abstract}
Layered \textit{Ln}OBiS$_2$ compounds with \textit{Ln} = La, Ce, Pr, Nd, and Yb can be rendered conducting and superconducting via two routes, substitution of F for O or the tetravalent ions Ti, Zr, Hf, and Th for trivalent \textit{Ln} ions.  Electrical resistivity measurements on non-fluorinated La$_{0.80}$Ti$_{0.20}$OBiS$_2$ and La$_{0.85}$Th$_{0.15}$OBiS$_2$ superconductors were performed between $\sim$1.5 K and 300 K and under pressure up to 2.4 GPa. For both compounds, the superconducting transition temperature $T_c$, which is $\sim$2.9 K at ambient pressure, gradually increases with pressure to 3.2-3.7 K at $\sim$1 GPa, above which it is suppressed and the superconducting transitions become very broad. Measurements of the normal state electrical resistivity of the two compounds reveal discontinuous changes of the resistivity as a function of pressure at $\sim$0.6 GPa, above which metallic-like electrical conduction is suppressed with increasing pressure.  Surprisingly, above 1.3 GPa, semiconducting-like behavior reappears in La$_{0.80}$Ti$_{0.20}$OBiS$_2$. This study reveals a new high-pressure phase of La$_{1-x}$$M$$_x$OBiS$_2$ containing the tetravalent ions $M$ = Ti, Th which does not favor superconductivity.  In contrast, application of pressure to fluorinated LaO$_0.5$F$_0.5$BiS$_2$ produces an abrupt tetragonal-monoclinic transition to a metallic phase with an enhanced $T_c$.  These results demonstrate that the response of the normal and superconducting properties of LaOBiS$_2$-based compounds depends strongly on the atomic site where the electron donor ions are substituted.  
\end{abstract}

\maketitle
keywords: BiS$_2$-based compounds; superconductivity; metal-semiconductor transition; pressure

\section{Introduction}

Following the discovery of superconductivity in the layered bismuth oxysulfide compound Bi$_4$O$_4$S$_3$, an intense effort to find new BiS$_2$-based superconductors quickly ensued as a result of the remarkable flexibility of the BiS$_2$-layers in forming compounds with new chemical compositions~\cite{mizuguchi2012bis2,yazici2013superconductivity,fang2015chemical,yazici2015superconductivity}. By doping the BiS$_2$-layers with electrons via appropriate chemical substitutions, superconductivity was found in \textit{Ln}(O, F)BiS$_2$ (\textit{Ln} = La-Nd, Yb), (Sr, La)FBiS$_2$, (La, \textit{M})OBiS$_2$ (\textit{M} = Ti, Zr, Hf, Th), Bi$_3$O$_2$S$_3$, La(O, F)BiSe$_2$, EuBiS$_2$F, and Eu$_3$Bi$_2$S$_4$F$_4$ with values of the superconducting transition temperature $T_c$ ranging from 2.7 to 10.6 K~\cite{yazici2013superconductivity,fang2015enhancement,jeon2014effect,lin2013superconductivity,yazici2013superconductivity,yu2013superconductivity,tanaka2014first,zhai2014possible,zhai2014anomalous}. One of the fascinating features in the behavior of several of the BiS$_2$-based superconductors, including \textit{Ln}O$_{0.5}$F$_{0.5}$BiS$_2$ (\textit{Ln} = La-Nd, Yb), (La,Sm)O$_{0.5}$F$_{0.5}$BiS$_2$, Sr$_{0.5}$\textit{R}$_{0.5}$FBiS$_2$ (\textit{R} = La-Pr, Sm), LaO$_{0.5}$F$_{0.5}$BiSe$_2$, and EuBiS$_2$F, is the pressure-induced enhancement of $T_c$ that occurs at about 0.5-2.2 GPa, revealing the existence of two distinct superconducting phases~\cite{fang2015pressure, wolowiec2013,fang2017upper,Jha,jha2015appearance,fujioka2014pressure,guo2015evidence}. This phenomenon is believed to be associated with a structural phase transition, which has been demonstrated in the case of LaO$_{0.5}$F$_{0.5}$BiS$_2$, from a tetragonal structure to a monoclinic structure with a sudden change in unit cell volume at an external pressure ($P$) of $\sim$0.5 GPa~\cite{tomita2014pressure}. 

In addition to the requisite superconducting BiS$_2$ layers, there are other similarities in the compounds reported to have a second superconducting phase under high pressure: 1) all of these compounds contain F, which resides in blocking layers and donates conduction electrons to the BiS$_2$ layers, and 2) the high pressure phase has a significantly larger value of $T_c$ than the low pressure phase. These similarities raise the issue of whether F is essential for the formation of the pressure-induced high-$T_c$ phase. Recently, superconductivity was also found in single crystalline CeOBiS$_2$ at 1.3 K, and the value of $T_c$ could be gradually enhanced under an external pressure~\cite{tanaka2017superconductivity}. This phenomenon was suggested to be related to Ce valence fluctuations, which can be tuned via the application of pressure; however, it is still not clear if there are concomitant phase transitions.  On the other hand, the compounds La$_{1-x}$$M$$_x$OBiS$_2$ ($M$ = Ti, Zr, Hf, Th) were reported to have the same crystal structure, very similar lattice parameters, and values of $T_c$ ($\sim$3 K) similar to that of LaO$_{0.5}$F$_{0.5}$BiS$_2$~\cite{yazici2013superconductivity2}. The $M$ ions in La$_{1-x}$$M$$_x$OBiS$_2$ are tetravalent and contribute conduction electrons to the BiS$_2$ layers in the absence of F; hence, this system is ideal for investigating whether non-F substituted BiS$_2$ compounds have a second pressure-induced high $T_c$ phase. High pressure studies of La$_{1-x}$$M$$_x$OBiS$_2$ also have the potential of uncovering new superconducting phases in BiS$_2$-based superconductors and will be very helpful in developing an understanding of the nature of the pressure-induced low-$T_c$ to high-$T_c$ phase transition in these materials.

\section{EXPERIMENTAL DETAILS}
Polycrystalline samples of La$_{0.80}$Ti$_{0.20}$OBiS$_2$ and La$_{0.85}$Th$_{0.15}$OBiS$_2$ were synthesized by solid state reaction as described elsewhere~\cite{yazici2013superconductivity2}. Geometric factors used in determining the electrical resistivity for each sample were measured before applying pressure. Hydrostatic pressure was generated by using a clamped piston-cylinder cell (PCC) in which a 1:1 mixture of n-pentane and isoamyl alcohol by volume was used as the pressure-transmitting medium. The pressures applied to the samples were estimated by measuring the $T_c$ of a high purity ($\textgreater$99.99\%) Sn disk inside the sample chamber of the cell and comparing the measured values with the well-determined $T_c$($P$) of high purity Sn~\cite{smith1969superconducting}. Electrical resistance measurements under high pressure up to $\sim$2.4 GPa were performed in a temperature range between 1.5 and 280 K using a standard four-probe method in a pumped $^4$He dewar.

\section{RESULTS AND DISCUSSION}

\begin{figure}[t]
\includegraphics[width=7.5 cm]{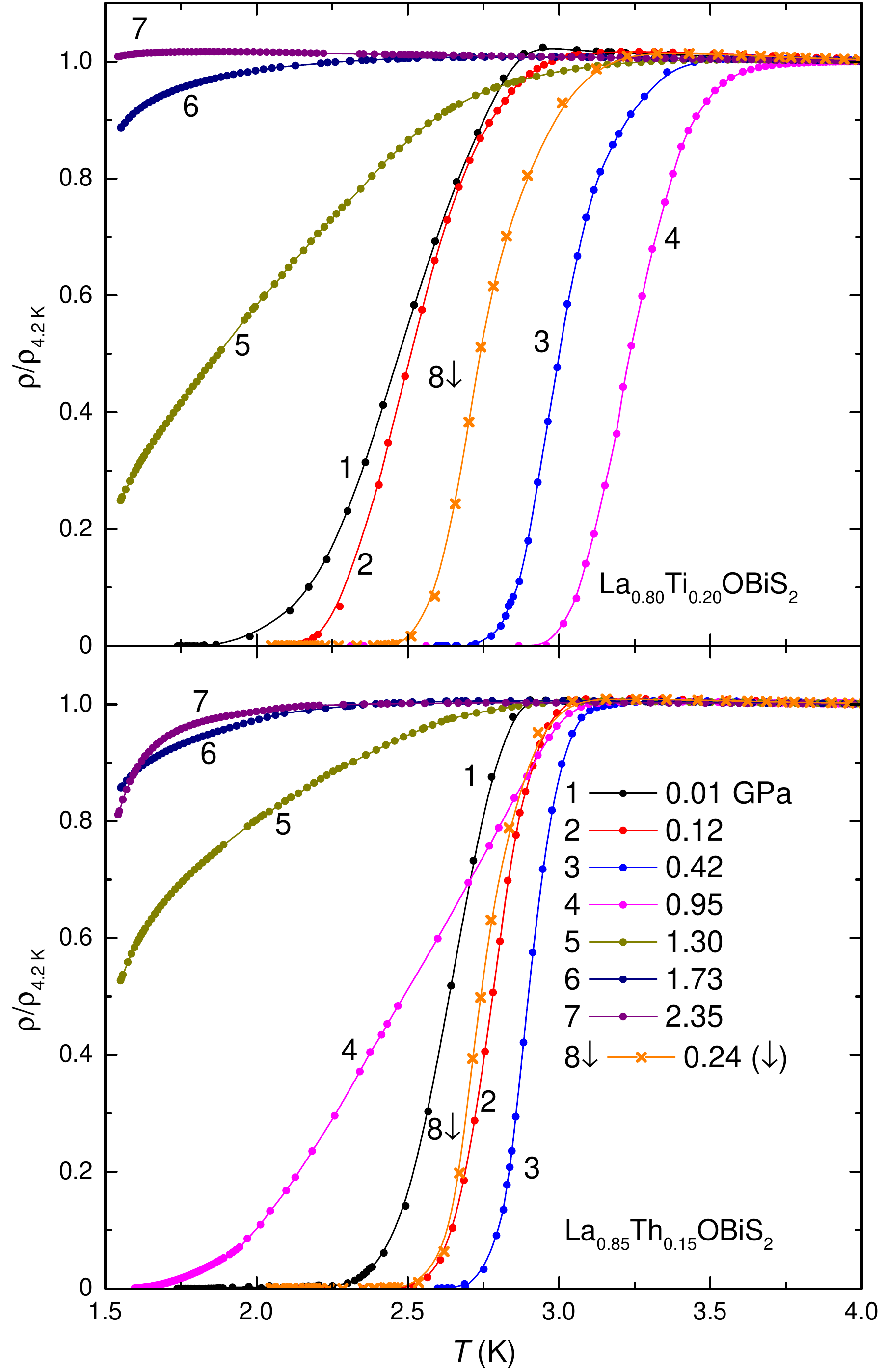}
\caption{(Color online) Temperature $T$ dependence of electrical resistivity $\rho$, normalized to its normal state value at 4.2 K, below 4 K for La$_{0.80}$Ti$_{0.20}$OBiS$_2$ (upper panel) and La$_{0.85}$Th$_{0.15}$OBiS$_2$ (lower panel). The numbers and symbols with different colors in both panels are used to indicate the pressures listed in the lower panel. Most of the $\rho$($T$) data (represented by circles) were obtained upon increasing pressure. Crosses in this figure and the following figures throughout this paper are used to represent data taken upon decreasing pressure.}
\label{FIG.1.}
\end{figure}

\begin{figure}[t]
\includegraphics[width=7.5 cm]{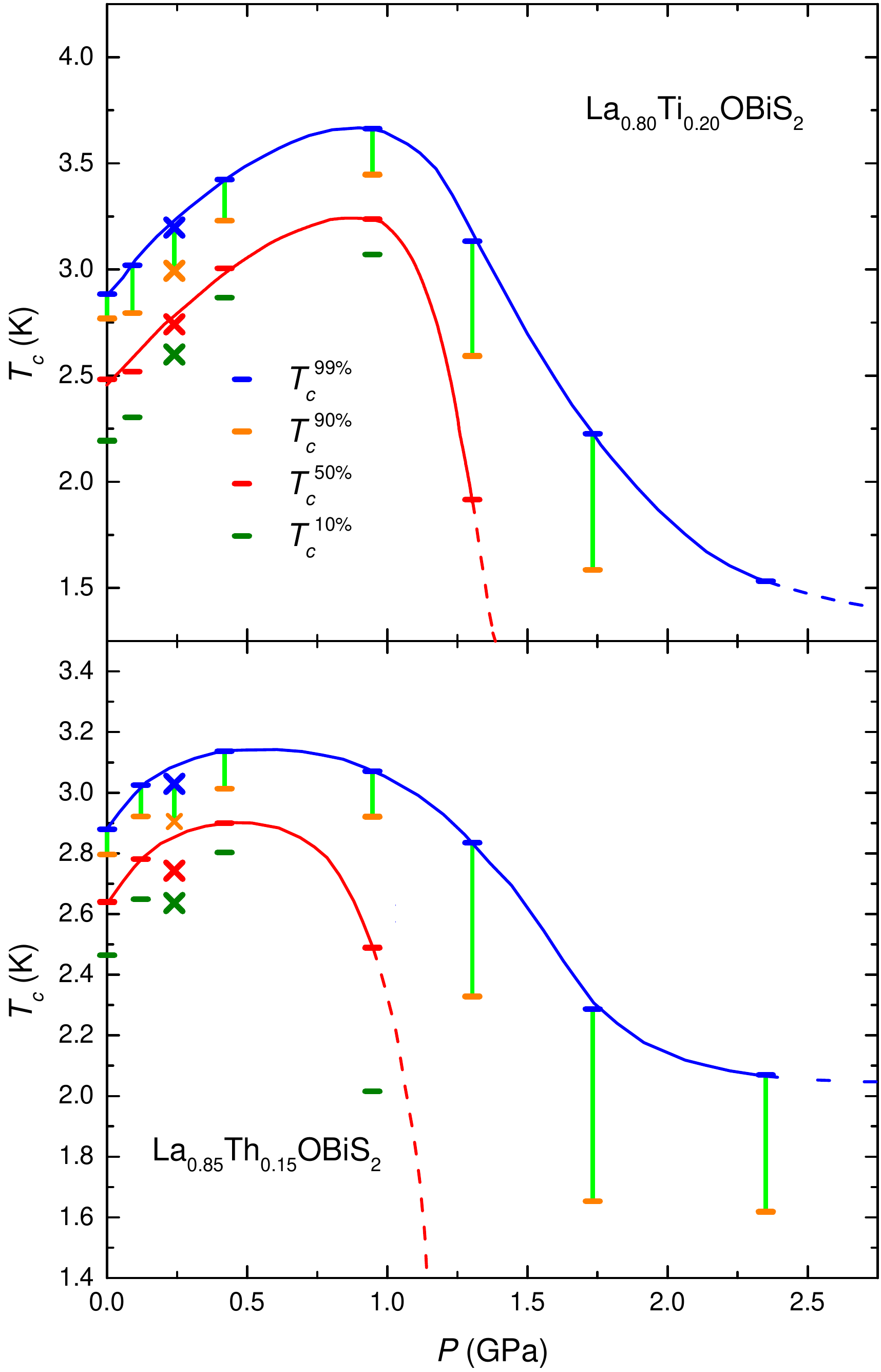}
\caption{(Color online) Superconducting transition temperature $T_c$ vs. pressure for La$_{0.80}$Ti$_{0.20}$OBiS$_2$ (upper panel) and La$_{0.85}$Th$_{0.15}$OBiS$_2$ (lower panel). The temperature at which $\rho$ drops to 99$\%$, 90$\%$, 50$\%$, and 10$\%$ of its normal state value above the superconducting transition is indicated by symbols $T_{c}^{99\%}$, $T_{c}^{90\%}$, $T_{c}^{50\%}$, and $T_{c}^{10\%}$, respectively. The length of the vertical lines represents the difference between $T_{c}^{99\%}$ and $T_{c}^{90\%}$ and is a measure of the broadening of superconducting transition under pressure.  Crosses represent data taken upon decreasing pressure. The solid and dashed lines are guides to the eye for the experimental data and the corresponding extrapolations, respectively.}
\label{FIG.1.}
\end{figure}

The resistive superconducting transition curves $\rho$($T$), normalized to the value of $\rho$ in the normal state at 4.2 K, for polycrystalline La$_{0.80}$Ti$_{0.20}$OBiS$_2$ and La$_{0.85}$Th$_{0.15}$OBiS$_2$ samples are shown in the upper and lower panels of Fig. 1, respectively. The superconducting transitions are very sharp for pressures below 0.95 GPa for both samples; however, the transitions become quite broad at higher pressures and the values of $\rho$ remain finite down to 1.5 K. The relationship between $T_c$ and pressure for the two compounds can be seen in Fig. 2. $T_c$$^{99\%}$, $T_c$$^{90\%}$, $T_c$$^{50\%}$, and $T_c$$^{10\%}$ represent the temperatures where $\rho$ falls to 99$\%$, 90$\%$, 50$\%$, and 10$\%$ of its normal-state value, respectively. The value of $T_c$$^{99\%}$ is 3.5 K for La$_{0.80}$Ti$_{0.20}$OBiS$_2$ at ambient pressure and increases slightly with increasing pressure to 3.7 K at 0.95 GPa, above which $T_c$$^{99\%}$ starts decreasing gradually. A similar dome-shaped $T_c$ vs. $P$ superconducting phase boundary can also be observed in the case of La$_{0.85}$Th$_{0.15}$OBiS$_2$, although the values of $T_c$ are slightly different than those of La$_{0.80}$Ti$_{0.20}$OBiS$_2$. For both compounds, the difference between $T_c$$^{99\%}$ and $T_c$$^{90\%}$ (indicated by the green vertical lines in Fig. 2), which is related to the width of superconducting transition, becomes much larger at pressures at which $T_c$ starts decreasing (0.95 GPa), reflecting dramatic changes in the superconductivity. During the pressure releasing cycle, the values of $T_c$ are reversible and the superconducting transitions become sharp again when the samples have returned to low pressures as can be seen in Figs. 1 and 2 (data represented by crosses). These features in the  $T_c$ vs. $P$ data indicate that there are significant differences in the superconducting behavior below and above 0.95 GPa.

To obtain further information about the origin of the suppression of superconductivity at high pressure, we investigated the normal state resistivity of the two compounds under pressure.  Upon warming, $\rho$($T$) of La$_{0.85}$Th$_{0.15}$OBiS$_2$ first decreases significantly, passes through a minimum, and then increases nearly linearly at higher temperatures, as shown in Fig.3. Such minima in the normal state resistivity are also observed for La$_{0.80}$Ti$_{0.20}$OBiS$_2$ under pressure and other BiS$_2$-based superconductors~\cite{fang2015pressure,yazici2013superconductivity2, demura2013new, zhai2014possible,kim2015observation}. In addition, subtle anomalies in resistivity, indicated by black arrows in Fig. 3, were observed in both compounds when pressure was applied to the samples. Upon further increase in pressure, the temperature at which the anomaly displays a kink ($T_a$) increases with increasing pressure (see the lower inset of Fig. 3); however, at pressures above 0.42 GPa, such anomalies in $\rho$($T$) can no longer be observed. Very recently, a similar kink in $\rho$($T$) was reported in non-superconducting LaOBiSe$_2$ at ambient pressure, which may related to a charge density wave transition and can be suppressed by F substitution~\cite{wu2015synthesis}. In this study, the appearance of these anomalies is possibly due to a temperature-driven phase transition; however, the nature of the transition will require further investigation.
%
%
%
%
  
\begin{figure}[t]
\includegraphics[width=7.5 cm]{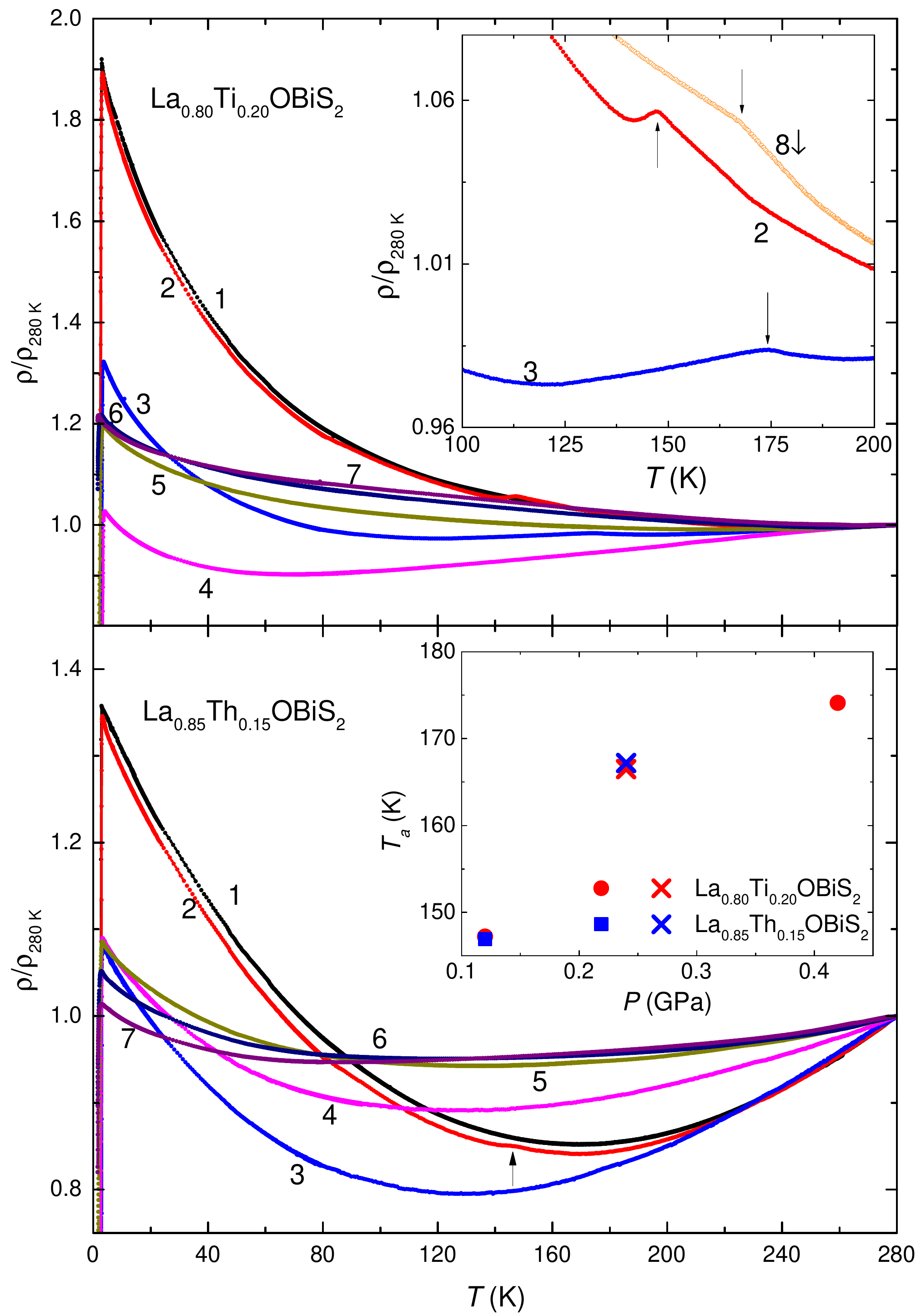}
\caption{(Color online) Temperature $T$ dependence of the electrical resistivity $\rho$, normalized to the value of $\rho$ at 280 K, ($\rho$/$\rho$$_{280}{\rm_{K}}$), of La$_{0.80}$Ti$_{0.20}$OBiS$_2$ (upper panel) and La$_{0.85}$Th$_{0.15}$OBiS$_2$ (lower panel) at various pressures. The pressure corresponding to each ($\rho$/$\rho$$_{280}{\rm_{K}}$) curve is indicated in Fig. 1 with the same color. Upper inset: enlargement of the $T$ dependence of $\rho$/$\rho$$_{280}{\rm_{K}}$ where the anomalies in $\rho$($T$) can be observed. The orange curve represents the data obtained at 0.24 GPa during the pressure releasing cycle. Lower inset: temperature at which the anomaly in $\rho$($T$) occurs, ($T_a$), vs. pressure ($P$). Black arrows in the figure indicate the location of anomalies in the resistivity.}
\label{FIG.3.}
\end{figure}

 In this work, we use the parameter $a$($T$), which represents the slope of $\rho$$(T)$, (d$\rho$/d$T$), at a certain temperature to characterize the tendency of the compound to exhibit metallic- or semiconducting-like behavior. For simple metals at high temperatures, $a$ is closely related to the scattering of conduction electrons by phonons. Although the value of $a$ does not have a clear physical significance for semiconductors, a positive value of $a$ suggests that the sample is metallic-like and a negative value of $a$ means it is semiconducting-like. In this work, the quantity $a$($T$) describes the tendency of the compounds to exhibit metallic- or semiconducting-like behavior at different pressures. As shown in Fig. 4(a), the values of $a$ at $\sim$10 K ($a_{10 \: \rm{K}}$) for the two compounds are negative at ambient pressure, suggesting semiconducting-like behavior at low temperatures. Upon application of an external pressures, $a_{10 \: \rm{K}}$ first increases rapidly at low pressures but relatively moderately at high pressures, resulting in a kink in the pressure dependence of $a_{10 \: \rm{K}}$ at $\sim$0.7 GPa. Although the values of $a_{10 \: \rm{K}}$ are all negative at pressures up to 2.4 GPa, the increase of $a_{10 \: \rm{K}}$ with pressure clearly indicates suppression of the semiconducting-like behavior. The kinks in $a_{10 \: \rm{K}}$($P$) found at $\sim$0.7 GPa suggests there are some fundamental differences between behavior of the two compounds at low and high pressure, which are consistent with the decrease of $T_c$ and broadening of the superconducting transitions at high pressure discussed previously.
  
 

Although the pressure dependences of $a$ at low temperature are quite similar for the two compounds, the values of $a$ at 250 K ($a_{250 \: \rm{K}}$) for La$_{0.85}$Th$_{0.15}$OBiS$_2$ are positive and almost the same at low pressures, in contrast to the remarkable enhancement of $a_{250 \: \rm{K}}$ from negative to positive values in the case of La$_{0.80}$Ti$_{0.20}$OBiS$_2$ as shown in Fig. 4(b). As indicated by the dashed lines in Fig. 4(b), the pressure dependence of $a_{250 \: \rm{K}}$ dramatically decreases at higher pressures for both compounds, which suggests not only pressure-induced suppression of metallic-like behavior at high temperatures for the two compounds but also indicates the appearance of high pressure phases that exhibit different types of electronic conduction compared to the corresponding low pressure phases. It should be noted that the values of $a_{250 \: \rm{K}}$ for La$_{0.80}$Ti$_{0.20}$OBiS$_2$ are even negative above 1.3 GPa, indicating the reappearance of semiconducting-like behavior.

 \begin{figure}[t]
 \includegraphics[width=7.5 cm]{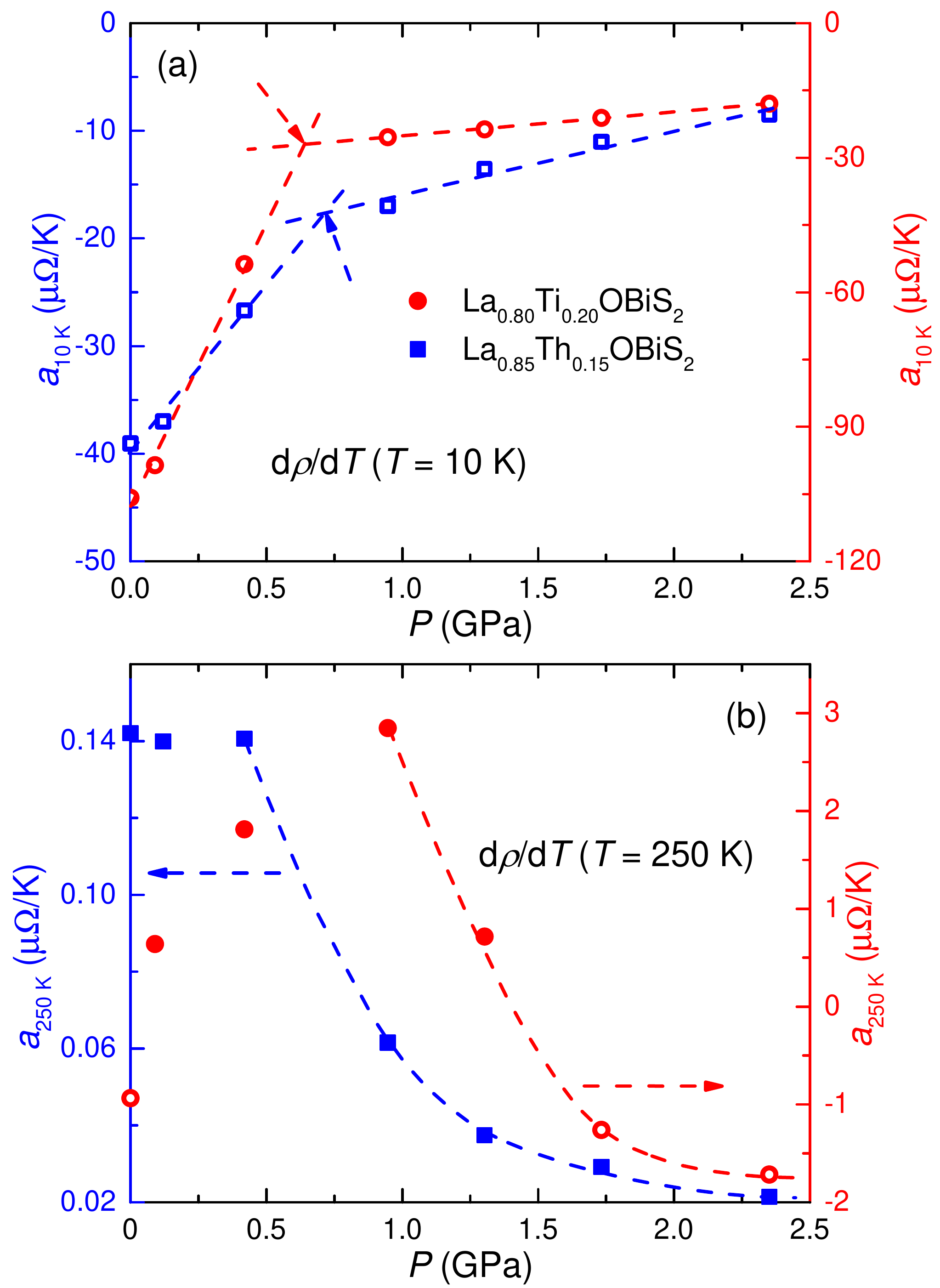}
 \caption{(Color online) Evolution of the slopes of $\rho$($T$) at $\sim$10 K (a) and at $\sim$250 K (b) with increasing pressure for La$_{0.80}$Ti$_{0.20}$OBiS$_2$ (red circles) and La$_{0.85}$Th$_{0.15}$OBiS$_2$ (blue squares), respectively. The open circles indicate that the values are negative, suggesting semiconducting-like behavior. Dashed lines are guides to the eye.}
 \label{FIG.4.}
 \end{figure}
 
Either the suppression of metallic conductivity or reappearance of semiconducting-like behavior is unusual for solids under pressure, since applying external pressures usually results in a significant increase in the carrier density. For a better understanding of the normal state resistivity, we plotted representative $\rho$ at 4.2 K ($\rho$$_{4.2 \: \rm{K}}$) vs. pressure data for the two compounds in Fig. 5. It can be observed that $\rho$$_{4.2 \:\rm{K}}$ for both compounds is significantly suppressed when pressure is first applied, which can be easily understood due to the increase of the electron density under pressures.  With further increase in pressure, however, $\rho$$_{4.2 \: \rm{K}}$ of La$_{0.80}$Ti$_{0.20}$OBiS$_2$ subtly increases and $\rho$$_{4.2 \: \rm{K}}$ of La$_{0.85}$Ti$_{0.15}$OBiS$_2$ decreases very slightly, resulting in kinks in $\rho$$_{4.2 \: \rm{K}}$($P$) at $\sim$0.6 GPa, as indicated by the arrows in Fig. 5. Consistent with previous discussions about the behavior of superconductivity and $a$ under pressure, a different phase whose normal state resistivity exhibits an unusual response to pressure appears at high pressures above $\sim$0.6 GPa.

\begin{figure}[t]
\includegraphics[width=7.5 cm]{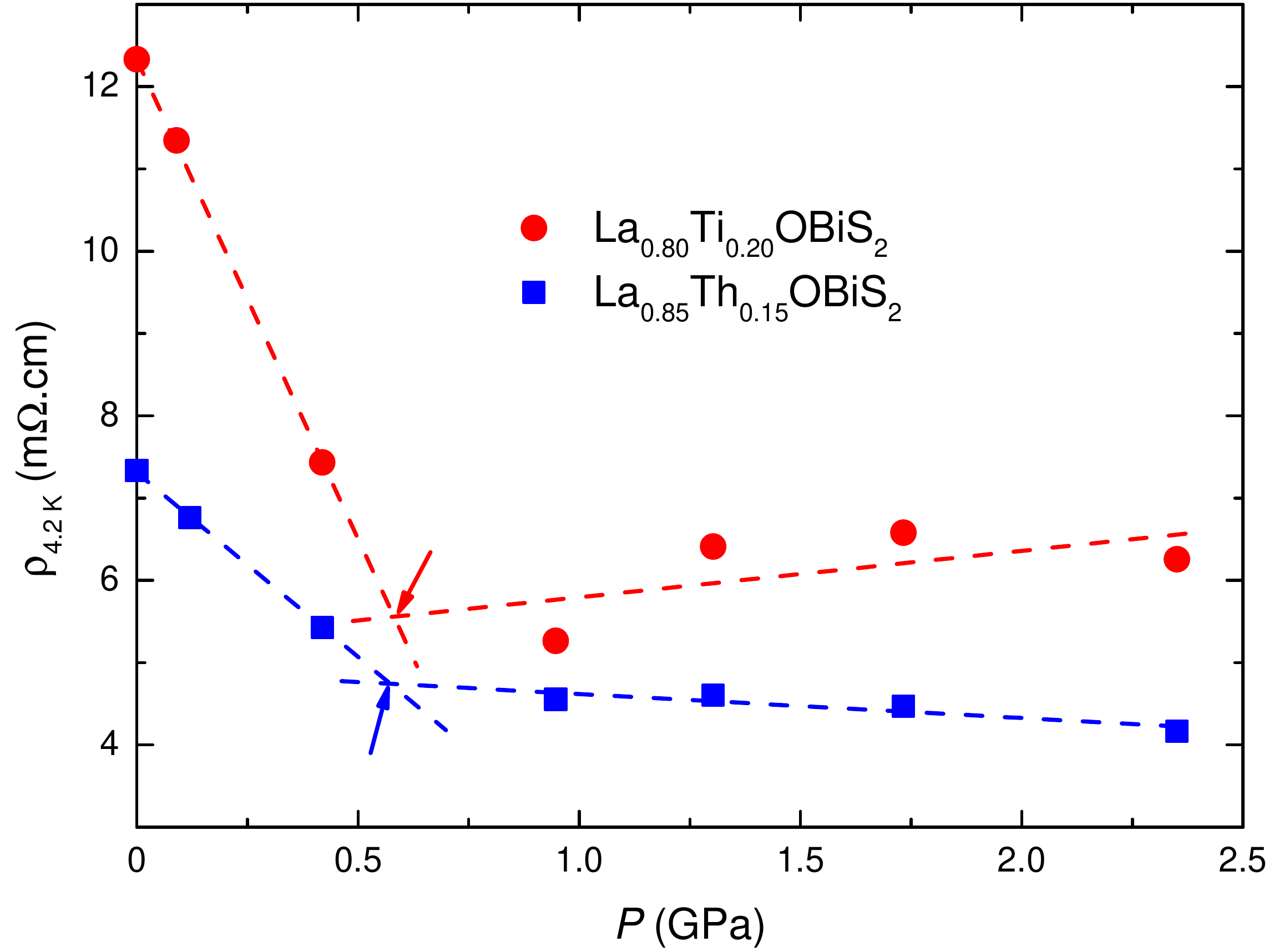}
\caption{(Color online) Normal state electrical resistivity $\rho$ at 4.2 K for La$_{0.80}$Ti$_{0.20}$OBiS$_2$ (red circles) and La$_{0.85}$Th$_{0.15}$OBiS$_2$ (blue squares) under pressure. The dashed lines are linear fits of $\rho$$_{4.2 \: \rm_{K}}$ vs. $P$ data whose slopes change at $\sim$0.6 GPa (indicated by the arrows) for both compounds.}
\label{FIG.5.}
\end{figure}

As mentioned in the introduction, all of the BiS$_2$-based superconductors which were reported to show pressure-induced low-$T_c$ to high-$T_c$ phase transitions contain F. The previous study of Sm substituted LaO$_{0.5}$F$_{0.5}$BiS$_2$ suggested that the pressure dependences of $T_c$ in BiS$_2$-based superconductors are closely related to the lattice parameter $a$ rather than the chemical composition of the blocking layers~\cite{fang2015pressure}.  In this study, although phase transitions were clearly observed in La$_{0.80}$Ti$_{0.20}$OBiS$_2$ and La$_{0.85}$Th$_{0.15}$OBiS$_2$ via investigations of the behavior of both the superconducting transitions and normal state resistivity under pressure, the high pressure phases of the two compounds are apparently unfavorable for superconductivity. The results indicate that F in the blocking layer of BiS$_2$-based compounds plays a critical role of not only providing conduction electrons but also the sudden pressure-induced enhancement of superconductivity.  This is presumably associated with a structural phase transition from tetragronal to orthorhombic symmetry, as has been shown for LaO$_{0.5}$F$_{0.5}$BiS$_2$. 

One may notice that although superconductivity is suppressed and $\rho$ remains finite to 1.5 K, the decrease of $\rho$ signaling the appearance of superconducting currents can still be observed up to the highest pressure obtained in this study. For homogeneous samples, in general, pressure-induced phase transitions (either structural or electronic, first or second order) usually occur in a rather narrow hydrostatic pressure range. The abrupt changes in normal state resistivity presented in this study also suggest the phase transition should be complete before 1 GPa; this raises the question of why the superconducting transitions of the two compounds remain broad at pressures above 1 GPa. It is possible that the high pressure phase is not superconducting and the superconducting currents observed above $\sim$1 GPa are associated with small regions of the low-pressure phase that remain at high pressure due to slight inhomogeneities in the samples. However, the possibility that the high pressure state of the compounds is still superconducting at lower temperature cannot be ruled out. Further studies need to be performed to determine whether the compounds are superconducting at lower temperatures.

\section{Concluding remarks}

In summary, the observed phenomena including suppression and broadening of the superconducting transition, changes of the normal state resistivity with pressure, and the dramatic changes in the normal state electron conduction behavior between 0.42-0.95 GPa strongly suggest a pressure-induced phase transition in the La$_{1-x}$$M$$_x$OBiS$_2$ compounds. In contrast to the pressure-induced low-$T_c$ to high $T_c$ phase transitions reported in F-containing BiS$_2$ superconductors, the high-pressure phase observed in this work does not favor superconductivity.  The results suggest that fluorine plays a critical role for the pressure-induced enhancement of $T_c$ for BiS$_2$-based compounds. For the low-pressure phase, anomalies in the normal state resistivity, which seem to be pressure dependent, were observed in the pressure range from 0.24 to 0.45 GPa. Another unusual phenomenon observed in this study is the tendency toward more semiconducting behavior with increasing pressure in the high-pressure phase. Further studies of these unusual phenomena are expected to enrich our understanding of the physics of solids at high-pressure.

\begin{acknowledgements}

High pressure research at UCSD was supported by the National Nuclear Security Administration under the Stewardship Science Academic Alliance Program through the US Department of Energy under Grant No.~DE-NA0002909.  Materials synthesis and characterization at UCSD was supported by the US Department of Energy, Office of Basic Energy Sciences, Division of Materials Sciences and Engineering, under Grant No.~DEFG02-04-ER46105.  Low temperature measurements at UCSD were sponsored by the National Science Foundation under Grant No.~DMR 1206553.

\end{acknowledgements}

\bibliography{endnote2}

\clearpage

\end{document}